\documentclass[aps,pre,onecolumn,superscriptaddress,showpacs,nofootinbib]{revtex4-2}
\usepackage{graphicx,svg,amsbsy,amsmath,epsfig,float}
\usepackage{amssymb,latexsym,wasysym,pifont,mathrsfs,gensymb}
\usepackage{comment,verbatim,multirow,array,sidecap,color}
\usepackage{lineno,setspace,soul,cancel,enumerate,amsmath,mathtools}
\usepackage[normalem]{ulem}
\usepackage{bm,bbm}

\begin{document}

\title{Fatigue failure in two-dimensional glasses under cyclic shear deformation}

\author{Himangsu Bhaumik}
\affiliation{Jawaharlal Nehru Centre for Advanced Scientific Research, Jakkur Campus,  Bengaluru, 560064, India}
\author{Srikanth Sastry}
\email{Author to whom any correspondence should be addressed: sastry@jncasr.ac.in}
\affiliation{Jawaharlal Nehru Centre for Advanced Scientific Research, Jakkur Campus,  Bengaluru, 560064, India}

%\keywords{glass, fatigue failure, cyclic shear}

\begin{abstract}
We investigate fatigue failure in two-dimensional (2D) model glasses under cyclic shear deformation using atomistic simulations. We find that the number of cycles to failure diverges as a power law as the strain amplitude approaches the fatigue limit, with an exponent close to $-1$, in contrast to the exponent of $-2$ reported in three dimensions (3D). A failure exponent of $-1$ has also recently been observed in a 2D  elastoplastic model, suggesting an interesting dependence on spatial dimensionality that needs to be rationalized. Measures of accumulated plastic activity, including dissipated work and non-affine displacements, exhibit scaling with the failure time that is consistent with results in 3D, and indicate a robust connection between damage accumulation and failure. To probe the origins of the variability of failure times, we perform isoconfigurational {\it seeded} simulations in which a localized soft region is introduced. While such seeding constrains the spatial location of failure, the distribution of failure times remains broad. These findings extend recent results in 3D concerning fatigue failure times to 2D glasses, including their relation to accumulated plasticity and apparent stochasticity, while also revealing a key difference, namely the exponent describing the divergence of the failure times. 
\end{abstract}

\maketitle
%\keywords{Suggested keywords}%Use showkeys class option if keyword                          %display desired

%\hbcom{EMP results for 2D has already been told in Pushkar's paper. If there is not 3D EPM, then we need to think whether we talk about EPM here or not.}

\section{Introduction}

Fatigue failure --- the eventual failure of materials subjected to repeated loading --- is a ubiquitous phenomenon in solids and is of central importance in materials science and engineering \cite{SureshCUP1998,schijve2009fatigue,basquin1910fatigue}. While fatigue in crystalline materials is often understood in terms of defect nucleation and crack growth \cite{SureshCUP1998,schijve2009fatigue}, the microscopic mechanisms underlying fatigue failure in amorphous solids remain less well understood because plastic deformation occurs through spatially heterogeneous and intermittent rearrangements rather than well-defined defects. Such plastic rearrangements in amorphous solids are commonly described in terms of localized shear transformations and interaction through long-range Eshelby stress fields \cite{argon1979plasticity,eshelby1957elastic,spaepen1977microscopic,picard2004elastic,FalkPRE1998,NicolasRevModPhys2018}.

The theoretical and computational investigation of yielding behaviour under cyclic shear deformation has attracted significant attention in recent years \cite{FioccoPRE2013,RegevPRE2013,Priezjev2013,LeishanthemNatCom2017,kawasakiPRE16,ParmarPRX2019,BhaumikPNAS2021,YehPRL2020,SastryPRL2021,MunganPRL2021,LiuJCP2022,ParleyPRL2022,CochranPRL2024,PRIEZJEV2021yielding,SarkarArxiv25,khandare2026,tiwari2026Pressure, ChatterjeeNATCOM2026}. Glasses subjected to cyclic shear strain undergo a yielding transition when the strain amplitude reaches a critical value, which we will refer to as the yield point. This transition is characterized by a change in the cycle-to-cycle particle dynamics—from non-diffusive to diffusive behaviour \cite{FioccoPRE2013,RegevPRE2013,FioccoIOP2015,LeishanthemNatCom2017,kawasakiPRE16,BhaumikJCP2022}, and is accompanied by strain localization and shear band formation beyond the transition \cite{ParmarPRX2019,BhaumikPRL2022}.

The nature of this yielding transition depends strongly on the degree of annealing of the glass. Poorly annealed glasses undergo mechanically induced annealing, with their energies decreasing as the strain amplitude increases, up to the yield point, beyond which, the energies begin to increase \cite{LeishanthemNatCom2017}. In contrast, well-annealed glasses show minimal changes in their properties prior to yielding, followed by a discontinuous transition. The extent of discontinuity increases with the degree of annealing. In simulation investigations of glasses  prepared {\it via} instantaneous quenches from liquid configurations equilibrated at a specified temperature (often termed the parent temperature), such a qualitative change in behaviour has been observed near the mode-coupling temperature \cite{YehPRL2020,BhaumikPNAS2021}. 

The number of cycles of deformation before failure, the fatigue failure time, depends strongly on the amplitude of deformation above the yield value, and has recently been investigated in the case of cyclic shear as well as other deformation modes, through computer simulations, as well as through theoretical models \cite{Bhowmik_2022,BhowmikPRE2022,KurotaniCommunMat2022,MaityNatPhys2026,SarkarArxiv25,khandare2026}.
%\sscom{Please update citation list to include EPM, mean field theory, etc.\hbcom{added Sarkar, Khandare Arxiv version}} 
In particular, the computational studies of model glasses subjected to cyclic shear deformation have revealed a number of striking features of fatigue failure. It has been shown that the number of cycles to failure diverges as a power law upon approaching a fatigue limit defined by the yield strain amplitude \cite{FioccoPRE2013, LeishanthemNatCom2017, KurotaniCommunMat2022, Bhowmik_2022, PRIEZJEV2023112230, MaityNatPhys2026}. 
%\sscom{Here again, our paper is not the first one to discuss, so add a comprehensive citation list for cyclic shear}
Furthermore, measures of \textit{accumulated damage}, quantified through non-affine displacements or dissipated energy, exhibit striking relationships with the failure time, indicating that fatigue failure can be understood in terms of progressive damage accumulation associated with plastic rearrangements. The spatial organization of these rearrangements, including the percolation of regions of plastic activity prior to shear band formation, provides a microscopic picture linking local plasticity to macroscopic failure \cite{MaityNatPhys2026}. The observed relationship between accumulated plasticity and failure times enable accurate prediction of failure times based on plasticity quantification. However, failure times are observed to be broadly distributed. Although a plausible analysis in terms of multiplicative accumulation of damage offers a useful approach \cite{maity2026stochasticityfatiguefailuretimes}, a satisfactory elucidation of the mechanism leading to fatigue failure remains to be worked out (see \cite{SarkarArxiv25} for a recently explored approach, which however does not incorporate spatial heterogeneity). 

A recently developed mesoscopic elastoplastic model that incorporates energy-landscape descriptions of local yielding \cite{khandare2026} reproduces key features of cyclic deformation and fatigue, including the divergence of failure times near yielding. Notably, in two dimensions, this model predicts a failure-time exponent of  $\approx -1$ (compared to $-2$, observed in 3D simulations \cite{MaityNatPhys2026}), while exhibiting similar relationships between accumulated damage and failure time.

Since the elastoplastic model investigated is in 2D, a natural question to address is whether the observed difference in the failure exponent value arises from the difference in the spatial dimensionality. Although not very well investigated, previous studies of yielding and plastic flow have shown that avalanche statistics, localization behaviour, and the nature of the yielding transition have dependences on spatial dimensionality \cite{lin2014scaling,budrikis2013avalanche,BhaumikPRL2022,salerno2013effect,NicolasRevModPhys2018,BhaumikJCP2022}. Such a role of spatial dimensionality is also of interest to investigate as an approach to developing insight into the process of damage accumulation and propagation that results in fatigue failure. 

%Plastic rearrangements in amorphous solids interact through long-range elastic fields whose spatial structure depends on dimension \cite{eshelby1957elastic,picard2004elastic}, and the geometry of localization and percolation of plastic activity is also expected to differ between two- and three-dimensional systems. Previous studies of yielding and plastic flow have shown that avalanche statistics, localization behaviour, and the nature of the yielding transition can depend sensitively on dimension \cite{lin2014scaling,budrikis2013avalanche,BhaumikPRL2022,salerno2013effect,NicolasRevModPhys2018,BhaumikJCP2022}. It is therefore natural to ask whether the recently identified scaling laws governing fatigue failure are universal or dimension-dependent.

In this work, we investigate fatigue failure in a two-dimensional model glass subjected to cyclic shear deformation, using atomistic simulations employed in earlier three-dimensional studies \cite{MaityNatPhys2026}. We consider two different models with substantially different interaction potentials in order to test the robustness of the observed behaviour. We focus on the dependence of the failure times on strain amplitude and on the relationship between failure times and measures of accumulated plasticity. We find that although the qualitative phenomenology of fatigue failure remains similar to that observed in three dimensions --- including the emergence of diffusive steady states, heterogeneous accumulation of plastic activity, and correlations between accumulated damage and failure times --- the exponent characterizing the divergence of failure times upon approaching the yield point from above is $-1$ in 2D, compared to $-2$ in 3D \cite{MaityNatPhys2026}. Our atomistic simulations thus suggest that the divergence exponent of $-1$ observed in the elastoplastic model investigation indeed arises from the spatial dimensionality of the model. In contrast, measures of accumulated damage based on dissipated work and non-affine displacements exhibit power-law relationships with failure times  with exponents that are close to to their three-dimensional counterparts.

%To test whether these results reflect generic collective plastic dynamics rather than atomistic details, we also investigate fatigue failure using a two-dimensional mesoscopic elastoplastic model (EPM) describing interacting shear transformations coupled through long-range elastic interactions \cite{picard2004elastic,jagla2010model,martens2011spontaneous,nicolas2018deformation}. The mesoscopic model reproduces the same qualitative fatigue phenomenology observed in atomistic simulations and yields a failure-time exponent close to $-1$, along with compatible scaling relations for accumulated plastic activity. The agreement between atomistic and mesoscopic descriptions provides strong evidence that the observed scaling behavior originates from the collective dynamics of plastic rearrangements in two dimensions.

Our findings highlight the importance of dimensionality in determining the kinetics of fatigue failure in amorphous solids and suggest that further investigations into such spatial dimensionality dependence may offer  new insights into how microscopic plastic rearrangements collectively lead to macroscopic failure.

%\sscom{Check J Phys A format and if required, add a paragraph here that describes how the rest of the paepr is organized.\hbcom{A template of JPhys A I have added in additional material folder.}}

The paper is organized as follows: In section 2 we describe the model and simulation details. In section 3 we describe the results. In section 4, we provide a summary and conclusions.

\section{Model and Simulation details:}
\subsection{2D LJ Model:}
We consider a binary Lennard-Jones mixture in two dimensions (2D LJ), introduced in Ref.~\cite{LanconJPF1984}, that has been widely used to investigate the mechanical properties of amorphous solids \cite{shi2005strain,ShiPRL07,BarbotPRE18,BarbotPRE2020}. The inter-atomic potential is given by
\begin{equation}
U_{ij}=\left\{
\begin{array}{lr}
4 \epsilon_{ij}[(\frac{\sigma_{ij}}{r_{ij}})^{12}-(\frac{\sigma_{ij}}{r_{ij}})^6]+A, &r_{ij} < R_{in}\\ \sum_{k=0}^{4}C_k(r_{ij}-R_{in})^k,&R_{in} < r_{ij}< R_{\rm{cut}}\\0,&r_{ij}>R_{cut}
\end{array}
\right.
\label{eqnpot}
\end{equation}
where the constants (A, C's) can be found in Ref. \cite{BhaumikJCP2022}. The density of the system is kept constant and equals $1.02$ in reduced units (described below).  The composition of the mixture has a number ratio of large ($L$) and small ($S$) particles, $N_L : N_S = (1 + \sqrt 5):4$. We use reduced units which are expressed in terms of the mass $m = m_S = m_L$, and the energy scale $\epsilon_{SL}$, length scale $\sigma_{SL}$ defining the inter species interaction. For this model $\sigma_{SS}/\sigma_{SL}=2\sin(\pi/10)$, $\sigma_{SS}/\sigma_{SL}=2\sin(\pi/5)$ and $\epsilon_{SS}/\epsilon_{SL}=\epsilon_{LL}/\epsilon_{SL}=0.5$. Time is expressed in  units of $\tau = \sigma_{SL} \sqrt{m/\epsilon_{SL}}$ and temperature is expressed in units of $\epsilon_{SL}/k_B$ where $k_B$ is the Boltzmann constant. The potential and its first two derivatives go continuously to zero at cutoff distance $R_{{cut}} = 2.5\sigma_{SL}$, and the potential is twice continuously differentiable at $R_{{in}} = 2\sigma_{SL}$. 

%\subsection{2D LJ 65:35 Model}
\subsection{2D Polydisperse Model:}We study a two-dimensional polydisperse (2D Poly) system of particles interacting {\it via} purely repulsive soft-disk interactions \cite{NinarelloPRX2017}. The particle diameters, $\sigma_i$, are sampled from a continuous distribution, $
f(\sigma) \propto \sigma^{-3}$, $\sigma \in [\sigma_{\min}, \sigma_{\max}]$ with the normalization constant fixed accordingly. The degree of size dispersity is characterized by $ \delta = \frac{\sqrt{\langle \sigma^2 \rangle - \langle \sigma \rangle^2}}{\langle \sigma \rangle}$. We set $\delta = 0.23$, by choosing $\sigma_{\min}/\sigma_{\max} = 0.45$. Lengths are expressed in units of the mean diameter $\langle \sigma \rangle $. Particles interact {\it via}  a pair potential of inverse power-law form, modified to ensure smooth truncation,
\begin{equation}
U_{ij}=  \epsilon_0\left [\left( \frac{\sigma_{ij}}{r} \right)^{12} + c_0 + c_1 \left( \frac{r}{\sigma_{ij}} \right)^2 + c_2 \left( \frac{r}{\sigma_{ij}} \right)^4\right ],
\end{equation}
where the effective interaction diameter is defined as
$ \sigma_{ij} = \frac{\sigma_i + \sigma_j}{2} \left(1 - \epsilon |\sigma_i - \sigma_j|\right)$. Here, $\epsilon$ is the degree of non-additivity of the particle diameter, and we choose $\epsilon = 0.2$ to improve glass-forming ability, following \cite{NinarelloPRX2017}. The coefficients $c_0$, $c_1$, and $c_2$ are determined such that the potential and its first two derivatives go to zero continuously at the cutoff distance $r_{\mathrm{cut}} = 1.25\,\sigma_{ij}$. For this model we choose the number density 
$1$ in reduced units. As in the case of the 2D LJ model, we use reduced units, with distance measured in terms of the mean particle diameter (denoted by $\sigma$ subsequently) and energy measured in units of $\epsilon_0$. 

%\sscom{Specify reduced units \hbcom{done minimally}.}

\subsection{Initial glass preparation:} 
\noindent{\bf 2D LJ:} We perform constant temperature and volume (NVT) molecular dynamics simulations using a Nos\'e--Hoover thermostat with an integration time step of $0.001$. The system is first equilibrated in the liquid state at two differnt temperatures $T=2.98$ and $0.35$ for a duration of $10^4\tau_\alpha$ and $100\tau_\alpha$, respectively, where $\tau_\alpha$ is the structural relaxation time. We obtain two sets of glasses from well-equilibrated high-temperature liquid (HTL, $T=2.98$) and supercooled liquid (ESL, $T=0.35$) configurations by performing local energy minimization to obtain the corresponding inherent structures, with $E_{IS}=-2.11$ and $-2.36$, respectively. More stable glasses (well-annealed) with lower $E_{IS}$ are generated using the finite-temperature and shear-rate annealing protocol of Ref. \cite{DasJCP22}, starting from ESL  configurations.
%referred to as ESL (Equilibrium Supercooled Liquid) glasses. 
%To generate more stable (well-annealed) glasses with lower inherent structure energies, we employ the finite-temperature and shear-rate annealing protocol described in Ref. \cite{DasJCP22}, starting from these IS configurations. 
This procedure allows the system to gradually explore lower-energy states. Using this protocol, we obtain glasses with $E_{IS} = -2.41$, which we refer to as WAL1 (Well Annealed Liquid), following Ref. \cite{BhaumikJCP2022}. Extrapolation of inherent structure energies from equilibrium simulations suggests that this energy corresponds to an effective parent temperature $T_p = 0.271$ \cite{BhaumikJCP2022}. Note that the mode coupling temperature ($T_{MCT}$) for this model has been reported to be approximately $0.37$ \cite{BarbotPRE2020}. Unless otherwise stated, all the results shown are for WAL1 glasses.

\noindent{\bf 2D Poly: } We employ the SWAP Monte Carlo algorithm, following Ref. \cite{NinarelloPRX2017}, to generate well-annealed configurations of the polydisperse glass. In each Monte Carlo step, a swap move---in which the diameters of a randomly selected pair of particles are exchanged---is attempted with probability $P_{\mathrm{swap}}$, and a translational move, describing a random displacement of a particle by an amount $\delta l$, is attempted with probability $1 - P_{\mathrm{swap}}$. Both moves are accepted according to the Metropolis criterion. We choose $P_{\mathrm{swap}} = 0.2$ and $\delta l = 0.1$. The system is equilibrated at a parent temperature $T_p = 0.062$. The equilibrium configurations are then energy-minimized to obtain the corresponding glassy states. Note that the mode coupling temperature ($T_{MCT}$) for this model has been reported to be approximately $0.11$ \cite{OzawaPRR2020}.

\subsection{Cyclic shear simulations:}
We perform non-equilibrium molecular dynamics simulations to implement  strain-controlled cyclic shear at finite temperatures and strain rates, by solving the Nos\'e-Hoover thermostatted SLLOD equations motion \cite{EvansPRA1984}. 
% \begin{align}
%     \dot{\mathbf{r}}_i&=\frac{\mathbf{p}_i}{m} + \dot{\gamma}_{xy}\mathbf{r}_{yi}\hat{i}\nonumber\\
%     \dot{\mathbf{p}}_i&=\frac{\mathbf{F}_i}{m} - \dot{\gamma}_{xy}\mathbf{p}_{yi}\hat{i}-\alpha(t)\mathbf{p}_i
% \end{align}
The strain $\gamma_{xy}$ of the system is varied as $\gamma_{xy}(t) = \gamma_{\max} \sin(\omega t)$, where $\gamma_{\max}$ is the maximum strain amplitude of the cycles. The shear rate at the start of the cycles  $\Dot{\gamma}_{xy}(t=0)=\gamma_{\max}\omega$ is chosen to be  $0.001$. We choose the thermostat temperature to be $T=0.001$. For cyclic shear, we choose integration time step 0.01 and all other parameters are chosen to be same as in \cite{MaityNatPhys2026}.

%The damping coefficient $\alpha(t)$ for the Nos\'e-Hoover thermostat is $\alpha(t)=p_\eta(t)/Q$, where $p_\eta$ is momentum for thermostat coordinate $\eta$ which evolves as:
%\begin{equation}
 %   \dot{p}_\eta=\sum_i^N\mathbf{p}_i^2/m-N_f k_B T
%\end{equation}
%where $N_f$ is the number of degree of freedoms. $Q$ is the mass of the thermostat which is associated with thermostat relaxation time $\tau$ as: $Q=N_f k_B T \tau^2$. We have chosen to be $\tau=1.0$. 

We consider a system size of $N=10000$ for both the models. For statistics consider $64$ independent samples for each strain amplitude $\gamma_{\max}$. MC SWAP is performed by in-house code and other numerical simulations including molecular dynamics, cyclic shear, and energy minimization via conjugate-gradient algorithm, are performed using the in-house modified version of LAMMPS \cite{LAMMPS}.

%\sscom{We should describe the quantities measured here: \hbcom{done}}

\subsection{Quantification of accumulated plasticity:}
We evaluate the failure times (which we always express in units of the number of cycles) by monitoring the potential energy per particle at the end of each cycle, and  non-affine displacements of particles between successive cycles (termed  $D^2_{\min,cyc}$), and the non-affine displacement with respect to the initial configuration (termed $D^2_{\min,init}$). $D^2_{\min}$ of a particle $i$ from time $t^\prime$ to $t$ is calculated as \cite{FalkPRE1998}:
\begin{eqnarray}
D^2_{\min}(i)(t,t-t^\prime)&\!=\!&\sum_j\sum_\mu  \biggl ( r^\mu_j(t)-r_i^\mu(t)-\sum_\nu(\delta_{\mu\nu}+\epsilon_{\mu\nu})  \left[r_j^\nu(t^\prime)-r_i^\nu(t^\prime) \right] \biggr) ^2 
\end{eqnarray}
where the indices $\mu$ and $\nu$ denote spatial coordinates and the index $j$ runs over the \textit{neighbour} atoms of the reference atom $i$. Here, the \textit{neighbour} of an atom is defined as a particle within a distance of $2.5\sigma_{SL}$ for 2D LJ model \cite{FalkPRE1998} and $2.5  \sigma $ for 2D Poly model\cite{SinghPRM2020}. The local strain $\epsilon_{\alpha\beta}$ which minimizes $D^2_{\min}$, is calculated as 
\begin{align}
   \epsilon_{\mu\nu}&=\sum_\lambda X_{\mu\lambda} Y_{\lambda\nu}^{-1} - \delta_{\mu\nu},
\end{align}
where
\begin{align}
    X_{\mu\nu}=&\sum_j\left[r^\mu_j(t)-r^\mu_i(t)\right]\left[r^\nu_j(t^\prime)-r^\nu_i(t^\prime)\right],\nonumber\\ 
    Y_{\mu\nu}=&\sum_j\left[r^\mu_j(t^\prime)-r^\mu_i(t^\prime)\right] \left[r^\nu_j(t^\prime)-r^\nu_i(t^\prime)\right].\nonumber
\end{align}

We also investigate the relationship between accumulated plasticity or damage, and failure times. To this end, we quantify damage by computing the dissipated work in each cycle, from the area of the loop enclosed by the stress-strain curves, as detailed in \cite{MaityNatPhys2026}. Adding the per cycle contributions up to a given cycle number $t$ is termed accumulated damage, $D^{acc}_t$. 

\section{Results}
\begin{figure*}[!t]
\includegraphics[width=1\textwidth]{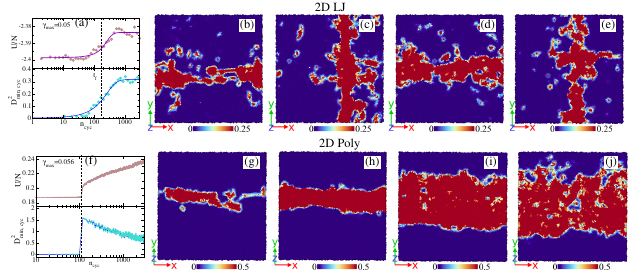}
    \caption{ Upper panels: 2D LJ model (WAL1), Lower panels: 2D Polydisperse Soft disk model. Evolution of energy (upper inset) and cycle-to-cycle non-affine displacement $D^2_{\min,cyc}$ (lower inset) as a function of strain cycle for (a) 2D LJ model (WAL1) and (f) 2D POly model. For the 2D LJ model, both $U/N$ and $D^2_{\min,cyc}$ can be well described by a sigmoid function as shown by the solid lines. Inflection points are taken to be the failure times. For the 2D Poly model, there is a sharp change in both $U/N$ and $D^2_{\min,cyc}$ during failure. The snapshots of the system coloured by the $D^2_{\min,cyc}$ of the particles are presented at different stages of evolution, (b,g) $n_{cyc}=t_f+10$, (c,h) $t_f+100$, (d,i) $t_f+1000$, and (e,j) $t_f+2500$ (from left to right) for both models. For 2D LJ $t_f=170$ and for 2D Poly $t_f=110$.
    }
\label{fig:fig1}
\end{figure*}

\subsection{Evolution of plastic activity and identification of failure}

We begin by characterizing the evolution of the system under cyclic shear deformation. Figure~\ref{fig:fig1} shows representative results for both the 2D LJ model and the 2D Polydisperse soft-disk model. The evolution of the potential energy per particle, $U/N$, and the cycle-to-cycle non-affine displacement $D^2_{\min,cyc}$\cite{FalkPRE1998} are shown as a function of the number of shear cycles. For the 2D LJ model, both $U/N$ and $D^2_{\min,cyc}$ exhibit a gradual transition from an initial stable regime to the yielded state. This evolution can be well described by a sigmoid fit as shown in the figure. The inflection point of the sigmoid fit provides a natural definition of the failure time $t_f$. For the 2D Poly model $U/N$ and $D^2_{\min,cyc}$ display a very sharp transition, with a rapid increase in both observables, indicating a sudden onset of irreversible plastic activity --- a brittle-like failure, as expected for a well annealed glass. For this system, the failure time $t_f$ is identified as the peak of the derivative of $U/N$ with respect to $n_{cyc}$. 

The spatial organization of plastic activity is illustrated by snapshots where particles are coloured with the $D^2_{\min,cyc}$ value (Fig.~\ref{fig:fig1}). In both models, we find that the plastic rearrangements are initially sparse and localized. However, at the failure point, the plastic rearrangements progressively organize into system-spanning structures -- similar to that observed for 3D glasses \cite{MaityNatPhys2026}. The appearance of a shear band indicates the macroscopic failure in the system. Interestingly, for the 2D LJ model, for the specific case shown, the shear band does not remain fixed in its orientation. Rather, its orientation alternates between the shear and gradient directions. The shear band for the 2D Poly system widens progressively while keeping its position and orientation fixed. We attribute the observed difference to the degree of annealing in the two cases, with 2D Poly configurations being better annealed, and note that shear banding in the shear and gradient directions has indeed been observed in other investigations \cite{ParmarPRX2019, GolkiaPRE2020,SinghPRM2020,OzawaPRR2020,FanPhysRevMaterials2022}.

%This apparent contrasting evolution of the shear band in these two models naturally raises the question about the statistics of the failure time in the two models, as well as how similar their failure mechanisms are -- which we will investigate next.

\subsection{Divergence of failure times near the fatigue limit}
\begin{figure}[!t]
\centerline{
\includegraphics[width=0.65\textwidth]{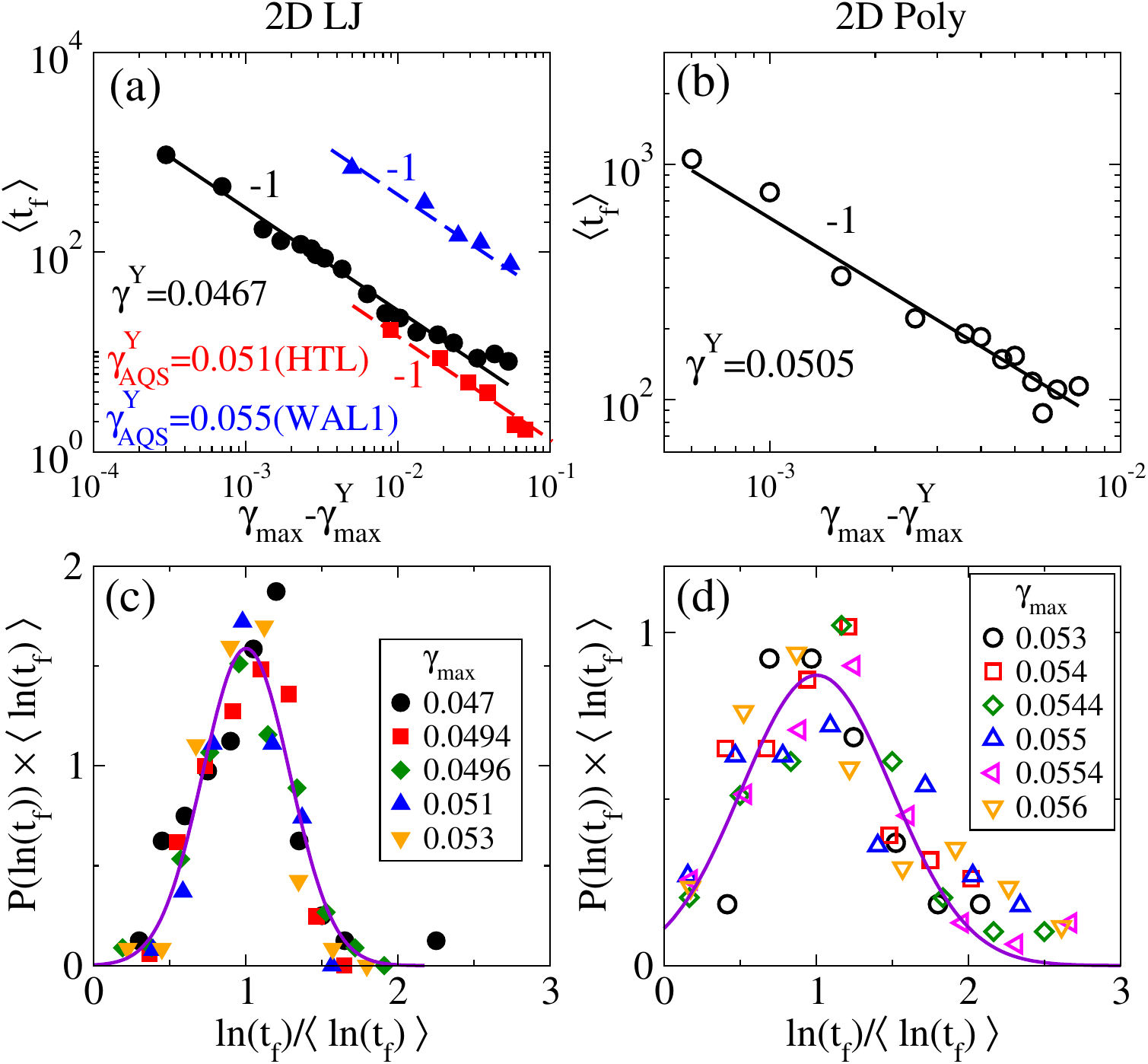}
}
\caption{Average failure times $\langle t_f \rangle$ against $(\gamma_{\max}-\gamma_{\max}^Y)$ for the (a) 2D LJ model and the (b) 2D Poly model. Solid lines are fits to a power-law which indicates that the exponent is $-1$, for both models. For comparison, results for the athermal quasistatic shear protocol are also shown for the 2D LJ model in (a), based on simulations reported in \cite{BhaumikJCP2022}. The distribution of failure times, for the (c) 2D LJ model and the (d) 2D Poly model, for several strain amplitudes. The distributions exhibit log-normal behaviour, with the width being proportional to the mean, as indicated by the data collapse when the axes are scaled by $\langle \ln (t_f) \rangle$.
}
\label{fig:fig2}
\end{figure}

We next investigate  the dependence of the sample-averaged failure time $\langle t_f \rangle$ on the strain amplitude $\gamma_{\max}$. Several forms of the dependence of the fatigue failure times, under different loading conditions and settings,  have been discussed in the literature \cite{SureshCUP1998,KunJStat2007,KunPRL2008,Bhowmik_2022}, but our recent study of three dimensional glasses \cite{MaityNatPhys2026} shows that the average failure time $\langle t_f \rangle$ diverges in a power-law fashion as the yield amplitude is approached from above. In Figure~\ref{fig:fig2}(a,c) we show the average failure time $\langle t_f \rangle$ as a function of the distance from the fatigue limit, $(\gamma_{\max} - \gamma_{\max}^Y)$, for the 2D LJ and 2D Poly models, respectively. In both cases, the failure time increases rapidly as $\gamma_{\max}$ approaches the yield strain $\gamma_{\max}^Y$ from above and is well described by a power-law divergence,
\begin{equation}
\langle t_f \rangle \sim (\gamma_{\max} - \gamma_{\max}^Y)^{\alpha},
\end{equation}
with an exponent $\alpha \approx -1$ \footnote{Free fits produce fit values of the exponent that are very close to $-1$ and we show fits in  Figure~\ref{fig:fig2} enforcing the exponent value to be $-1$.}. Notably, this exponent differs from the value $\alpha \approx -2$ reported in three-dimensional systems, indicating a clear dimensional dependence. 

For comparison, we also report in Figure~\ref{fig:fig2} (a) the failure times under the athermal quasistatic cyclic shear protocol for the 2D LJ model, based on results reported in Ref. \cite{BhaumikJCP2022}.  The the failure times are obtained from a stretched-exponential fit to the sample-averaged energy as a function of the strain cycle. As shown in Fig. \ref{fig:fig2}(a), both the poorly annealed (HTL, $E_{IS}=-2.11$) and well-annealed (WAL1, $E_{IS}=-2.41$) glasses exhibit a $\langle t_f \rangle \sim (\gamma_{\max}-\gamma_{\max}^Y)^{-1}$ scaling above the transition.

The distribution of failure times for different strain amplitudes is shown in Fig.~\ref{fig:fig2}(b,d). When the logarithm of the failure times are rescaled by their mean values, the distributions collapse onto a single master curve that is well described by a log-normal form. This is consistent with the observations for 3D glasses in Ref. \cite{maity2026stochasticityfatiguefailuretimes}, where it is argued that the form of the failure time distributions suggests that failure arises from a multiplicative stochastic process associated with the accumulation of plasticity over cycles. 

%\sscom{A log-normal does not suggest multiplicative process. text commented.}
%This suggests that failure arises from multiplicative stochastic processes associated with the accumulation of plastic activity over cycles.

\subsection{Relationship of accumulated plasticity with failure time}
\begin{figure}[!t]
\centerline{ 
\includegraphics[width=0.65\textwidth]{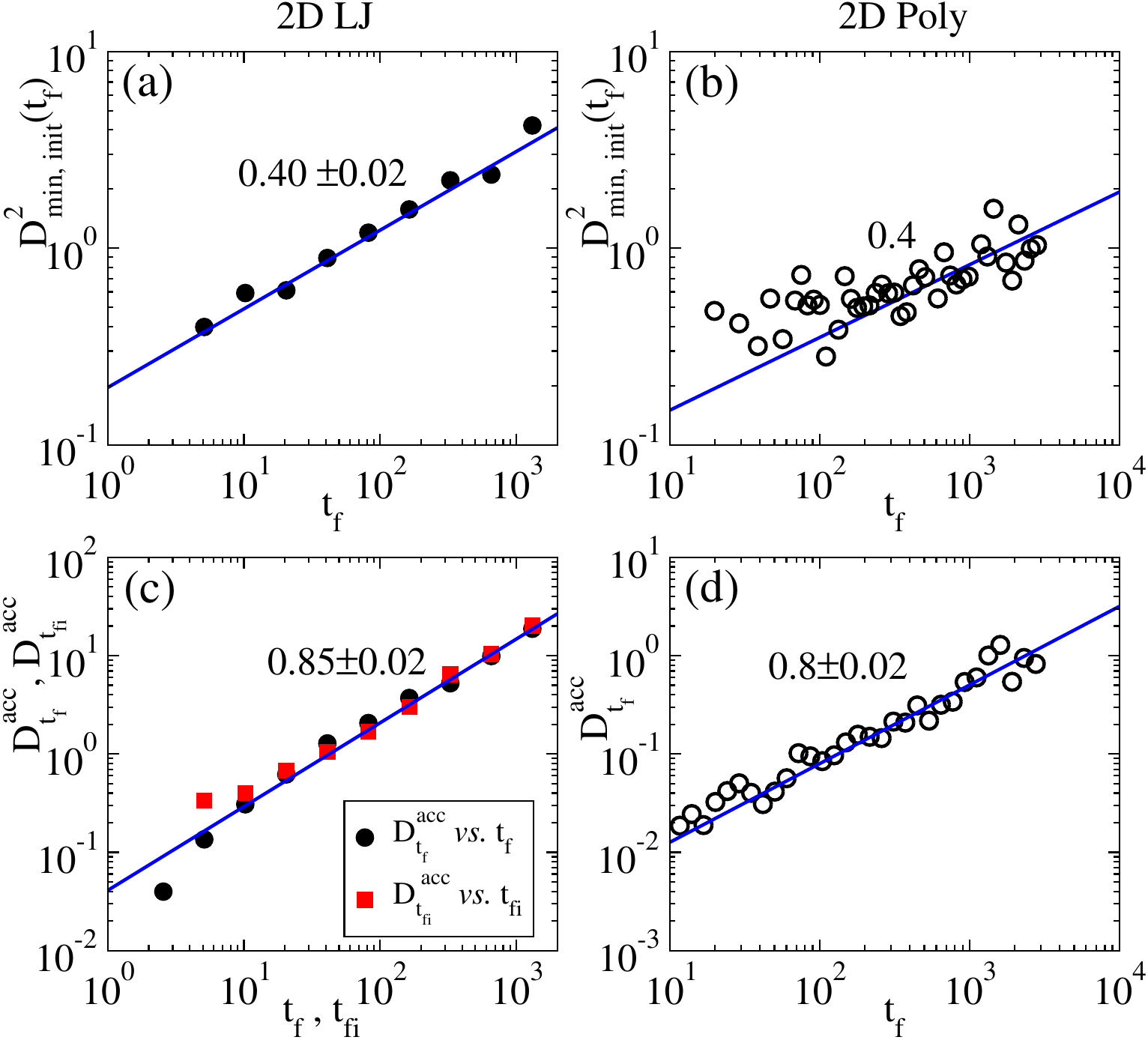}
}
    \caption{ Accumulated plasticity $D^2_{\min,init}$ (measured in terms of non-affine displacement) as a function of failure time $t_f$ for (a) 2D LJ and (b) 2D Poly. Solid lines through data represent power-law scaling   $D^2_{\min,init} \sim t_f^{0.4}$ for both models. Dissipated work $D^{acc}_{t_f}$ accumulated till failure as a function of failure time $t_f$ for (c) 2D LJ and (d) 2D Poly. A power-law $D^{acc}_{t_f} \sim t_f^{0.8}$ is observed for both the models. In (a), (c), and (d), the error bars on the exponent values represent the variation obtained by fitting different portions of the data. In (b), the line is a guide to the eye with slope $0.4$.
    }
\label{fig:fig3}
\end{figure}

\begin{figure*}[!t]
\centerline{
\includegraphics[width=1\textwidth]{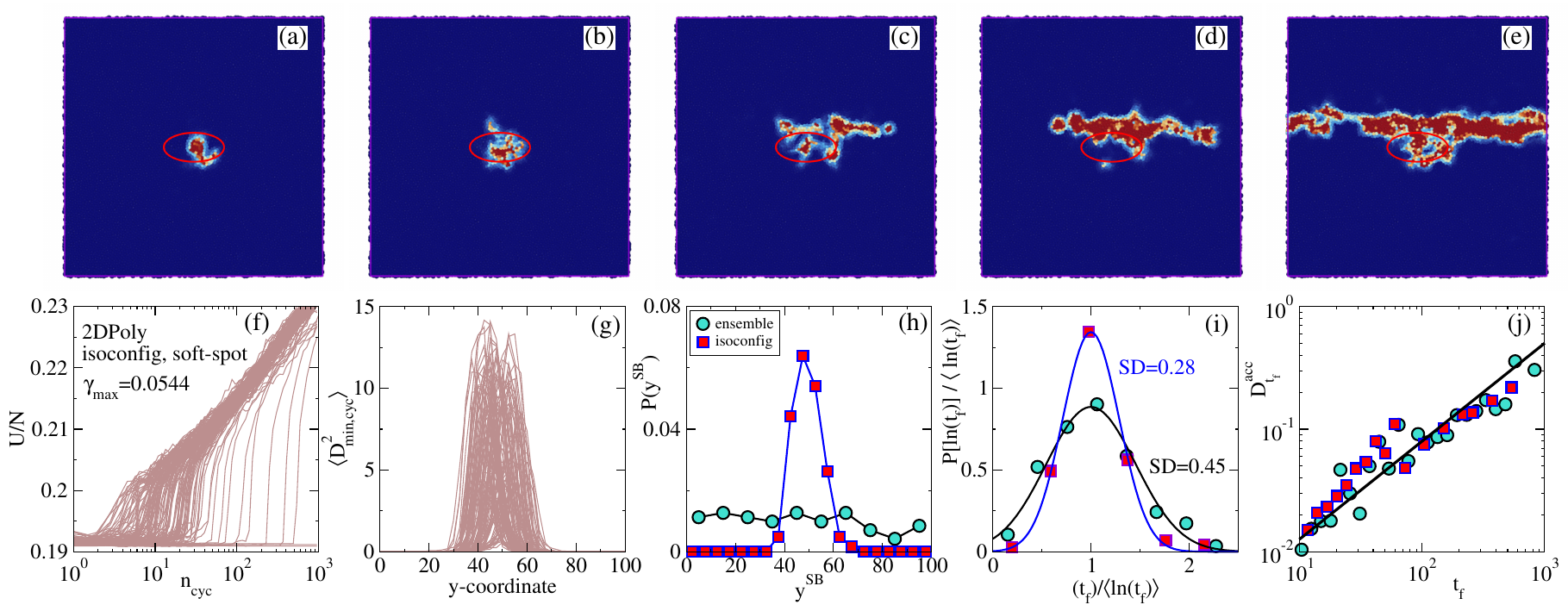}
}
\caption{Isoconfigurational simulations with a predefined soft spot (2D Poly). The upper panels (a)-(e) show the nucleation of the shear band from the seeded region and the subsequent evolution of plastic rearrangements up to failure. Lower panels: (f) Evolution of potential energy as a function of strain cycles for 200 independent runs with different initial velocities. (g) Cycle-to-cycle non-affine displacement profile at the failure time $t_f$, demonstrating that the shear band formation is spatially localized near the seeded region. (h) Distribution of shear band positions: while unseeded simulations exhibit a broad (nearly uniform) distribution, the seeded isoconfigurational runs show a pronounced peak at the location of the seed. (i) %\sscom{needs to be same colours.\hbcom{fixed}}
Distribution of failure times for both configurational ensemble and isoconfigurational runs. In both cases, the distributions remain broad and are well fitted by the log-normal distribution, although the seeded runs show a reduction in the standard deviation ($SD = 0.28$) over the unseeded runs ($SD = 0.45$). (j) Dissipated work as a function of failure time for both cases, exhibiting the same power-law scaling with exponent $\approx 0.8$. }
\label{fig:fig4}
\end{figure*}   

Having characterized the statistics of failure times, we next examine how plastic activity accumulates prior to failure, and whether it can provide any insight to the observed variability in $t_f$. In amorphous solids, plastic deformation proceeds through localized rearrangements, and the cumulative effect of such events is often associated with the notion of damage. However, the precise way in which such damage relates to failure remains unclear. Following Ref. \cite{MaityNatPhys2026}, we analyse different measures of damage: plastic activity through non-affine displacement and dissipated work.

%\sscom{Paragraphs are bolder than subsection headings. Fix. \hbcom{fixed}} 

\paragraph{Accumulated non-affine displacement.}
We first consider the non-affine displacement relative to the initial configuration, $D^2_{\min,init}$ \cite{FalkPRE1998}, which captures the cumulative deviation from the initial structure over successive cycles.
%The evolution of system averaged $D^2_{\mathrm{init}}$ with cycle number shows a gradual increase followed by a sharp rise in the vicinity of failure (not shown), consistent with the onset of large-scale plastic activity.
To investigate if the accumulated damage till the failure correlates with the failure time, we compute $D^2_{{\min,init}}$ at the failure time $t_f$ for different samples and strain amplitudes. The results are shown in Fig.~\ref{fig:fig3}(a),(b) for 2D LJ and 2D Poly respectively. We find that the data is well described by a power-law relation
\begin{equation}
D^2_{\min,init} \sim t_f^{\delta},
\end{equation}
with $\delta \approx 0.4$ for both models, although for 2D Poly, the data quality is poor and we do not attempt to obtain the exponent value by fitting. Instead, a power law with exponent $0.4$ is displayed in Fig.~\ref{fig:fig3} (b) along with the data to show that the data are consistent with such an exponent value. The observed power-law relationship, including the exponent value, are the same as what was previously observed for 3D glasses in Ref. \cite{MaityNatPhys2026}. Results in \cite{MaityNatPhys2026} considered the failure initiation time $t_{fi}$ rather than the failure time. We do not follow the same procedure here given the difficulty of distinguishing these times for the 2D Poly glasses, but the results apparently are not affected by this difference.

%The observed sub-linear dependence indicates that the accumulated non-affine displacement grows slowly with increasing failure time. In particular, samples that fail at later times do not accumulate proportionally larger early-stage damage, indicating that the rate of damage accumulation decreases with increasing lifetime. 

\paragraph{Accumulated dissipated work.}
We next consider the accumulated dissipated work as a measure of damage, and following previous work \cite{MaityNatPhys2026}, consider accumulated  dissipated work up to the time of failure, $D^{\mathrm{acc}}_{t_f}$. This quantity is calculated by integrating the stress-strain loop area for each cycle and summing them up to the failure cycle. As shown in Fig.~\ref{fig:fig3}(c),(d), this quantity exhibits a power-law dependence on the failure time, following
\begin{equation}
D^{\mathrm{acc}}_{t_f} \sim t_f^{\beta},
\end{equation}
with $\beta \approx 0.8$ for both the models we considered. The exponent $\beta < 1$ implies that the average damage accumulated per cycle decreases with increasing $t_f$, indicating that samples which fail later do so through slower effective accumulation of plastic activity. For 3D glasses, an exponent of $0.8$ was observed when the failure initiation time $t_{fi}$ was considered, whereas an exponent of $0.75$ was observed when the failure time $t_{f}$ was considered. Here, for the 2D LJ, a slightly larger exponent value of $0.85$ is observed {\it both} for with respect to $t_{fi}$ and $t_{f}$. On the other hand, the 2D Poly system exhibits an exponent value of $0.8$  with respect to failure time $t_{f}$. In addition to the 3D glasses, the 2D elastoplastic model investigated in  \cite{khandare2026} also exhibits an exponent value of $0.75$ with respect to $t_{f}$. Thus the 2D systems investigated here display exponent values that are close to but slightly higher than the 3D glasses and the 2D elastoplastic model. Whether the observed deviations arise from numerical uncertainties, or have significance, needs to be evaluated in future work, as indeed the significance of the nearly equal exponent values observed in 2D and 3D glasses. The origin of the relationships between the failure times and the studied measures of plasticity also require a theoretical rationalization.

%The fact that both $D^2_{\min,init}$ and $D^{\mathrm{acc}}_{t_f}$ exhibit systematic scaling with $t_f$, while not reaching a constant value at failure, indicates that failure is not associated with a sharply defined damage threshold. Instead, failure reflects a progressive accumulation process, whose variability is consistent with the stochastic nature of the underlying dynamics, similar to the three-dimensional glass studied before \cite{MaityNatPhys2026,NicolasRevModPhys2018}.

\subsection{Stochasticity in failure times with seeding.}
In addition to the divergence of failture time approaching the yield point, and the relationship with measures of plasticity, the distribution of failure times for 3D glasses and a 2D elastoplastic model have recently been investigated \cite{maity2026stochasticityfatiguefailuretimes}. It is found that failure times are broadly distributed, and are fit equally well by several distribution functions. Particularly, the log-normal and inverse Gaussian distributions describe the results equally well. The latter distribution can be interpreted as a result of a multiplicative stochastic process. In order to understand the variability of the failure times, a set of ``isoconfigurational'' cyclic shear runs were performed in \cite{maity2026stochasticityfatiguefailuretimes}, wherein for the same initial configurations, an independent realization of initial velocities was used for initializing the runs. Such isoconfigurational runs generate a distribution of failure times whose variance is smaller than, but comparable  to, the variance when an ensemble of initial configurations is used for initialization. 

We further investigate the origin of the variability of fatigue failure times using isoconfigurational simulations, but with {\it seeded} initial configurations for the 2D polydisperse model. In these simulations, a localized soft region is artificially introduced, following Refs.~\cite{OzawaPRR2022,ChaudhuriScPost2025}. Specifically, we select an elliptical region at the center of the system, with major and minor axes of $10$ and $5$ (in units of $\sigma$), respectively. We perform SWAP Monte Carlo for $10^6$ sweeps to thermalize the region  to a temperature $T_h = 5$, which is significantly higher than the mode-coupling temperature $T_{MCT} = 0.11$. The updates are restricted to the particles within this region, while particles outside the region are held fixed. The system is subsequently quenched to obtain a glass configuration containing an embedded soft spot. This procedure creates a localized region that is more susceptible to plastic rearrangements, thereby biasing the location of damage initiation. Consistent with this expectation, we observe that the shear band repeatedly emerges from the seeded region across independent runs, starting with the same initial seeded configuration, but different velocity realizations. 

The consequences of this controlled perturbation are summarized in Fig.~\ref{fig:fig4}. The upper panels show representative snapshots of the evolution of plastic activity. In all cases, plastic rearrangements nucleate within the seeded region and subsequently grow into a system-spanning shear band. This demonstrates that the spatial location of failure is effectively fixed by the imposed soft spot, in contrast to unseeded ensemble simulations where shear bands can form at arbitrary locations.

The lower panels quantify the statistics of this process. In Fig.~\ref{fig:fig4}(f), we plot the evolution of the potential energy as a function of cycle number for multiple ($200$) runs of the same seeded configuration with different initial velocities. While all trajectories exhibit a similar qualitative evolution and a sharp increase at failure, the cycle at which this occurs varies widely across runs, indicating persistent variability in failure times. In Fig.~\ref{fig:fig4}(g), we show the cycle-to-cycle non-affine displacement profile at the failure point. The profile exhibits a pronounced peak centred at the seeded region, confirming that plastic activity remains spatially localized and that the eventual shear band consistently originates from the imposed soft spot. This is further quantified in Fig.~\ref{fig:fig4}(h), which shows the distribution of shear band positions (along the y-axis). For unseeded simulations with an ensemble of ($200$) independent initial conditions, this distribution is uniform, reflecting the absence of a preferred failure location.

Despite this strong spatial bias, the temporal statistics remain largely unchanged. As shown in Fig.~\ref{fig:fig4}(i), the distribution of failure times for the seeded isoconfigurational runs remains broad, with a width that is comparable to that for the unseeded configurational ensemble, although with a slight reduction in variance. This indicates that fixing the location of failure does not eliminate the stochasticity in the failure times. Finally, Fig.~\ref{fig:fig4}(j) shows that the accumulated dissipated work up to failure continues to scale with the failure time in the same manner as in the unseeded case, following a power-law relation with exponent $0.8$. This further supports the conclusion that the underlying failure mechanism remains unchanged. These results demonstrate that while the spatial location of failure can be controlled by introducing a predefined soft spot, the stochasticity in failure time persists indicating the variability in fatigue failure is not solely  governed by structural disorder, but instead arises from the intrinsically stochastic nature of the dynamical evolution. 

\section{Summary and Conclusions}\label{sec:summary}
We have studied fatigue failure in two two-dimensional model glasses under cyclic shear. Although failure occurs with significantly different abruptness in these two systems, both systems share several common features. 
%In both systems, failure is preceded by the gradual build-up of plastic activity, which eventually leads to the formation of a system-spanning shear band. 
A key result of this work is that the failure time increases as a power law as the strain amplitude approaches the fatigue limit, with an exponent $\approx -1$, which is different than the 3D counterpart where the exponent is $-2$. Such dimensional dependence of the divergence times have not been reported earlier, to our knowledge, for related model or experimental systems. In two dimensions, the corresponding exponents have been investigated in the broader context of periodically driven systems \cite{ReichhardtPhysRevRes2023}, for sheared granular matter \cite{GhoshPRL2022},  a 2D model of sheared non-Brownian suspensions, the random organization model (ROM) \cite{CorteNATPHY2008}, and 2D stress controlled shear \cite{Goswami2025NatPhys}, reporting significantly different exponents ($-1.33$ on both sides of the transition for ROM \cite{CorteNATPHY2008}, $-3/4$ on both sides of the dynamical transition in experiments on cyclically sheared frictional monolayer disks \cite{GhoshPRL2022}, and $-1.87$ in stress controlled cyclic shear \cite{Goswami2025NatPhys}). The corresponding exponents in active driven systems show a wide range ($-1.27$ for particles with oscillating diameters \cite{tjhung2017pre}, and $\approx -4$ for 2D active Brownian particles \cite{Goswami2025NatPhys}). It will be of interest to have additional results for more model and experimental systems. It will also be interesting to develop a satisfactory rationalization of the dimensionality dependence of these exponent values. Dimensionality dependence in the context of yielding in glasses, and correspondingly related exponent values,  has been discussed in earlier work. Two-dimensional glasses have been found to exhibit stronger disorder-induced fluctuations and finite-size effects than their three-dimensional counterparts \cite{OzawaPRR2020}, while elastoplastic models yield different estimates of the correlation-length exponent in 2D and 3D \cite{RossiPRL2022}. These results suggest that enhanced spatial fluctuations in lower dimensions may contribute to the different fatigue exponents observed here, but further investigations will be needed to develop a meaningful explanation of the observations presented here.

Despite the difference in the failure-time exponent between 2D and 3D, the accumulated plastic activity exhibits scaling with the failure time for the two model glasses that is consistent with that observed in 3D models and in 2D elastoplastic model, suggesting that the connection between damage accumulation and failure time is robust across dimensions.

%\sscom{This needs to be moved to a better location. Also Regev et al don't look above the transition. You should make sure you are looking at the correct dimensionality, etc:  Related periodically driven systems have also shown a divergence of characteristic time scales near the onset of irreversible dynamics. Regev, Lookman, and Reichhardt \cite{RegevPRE2013} reported a power-law divergence of the relaxation time, with an exponent of approximately $-2.6$ in a two-dimensional model. Similarly, the random-organization model exhibits a power-law divergence of the time required to reach the steady state, with exponents of approximately $-1.33$ both below and above the transition in simulations, whereas in experiments it is $-1.1$ below the transition \cite{CorteNATPHY2008}, see also \cite{ReichhardtPhysRevRes2023}.

%So in the discussion section, summarize the comparisons we can make and any explanations that people may have offered for the difference with 3D.

%}

To better understand the stochastic nature of failure, we performed isoconfigurational simulations by introducing a unique localized soft region. This allows us to control where the shear band forms. We find that while the location of failure becomes fixed by the soft spot region, the failure time still varies significantly across runs. The distribution of failure times remains broad, similar to the unseeded case, but with a reduction in the variance. This shows that while the location of failure can be controlled, the timing of failure remains inherently stochastic. 

Overall, our results show that fatigue failure in two-dimensional glasses is governed by the progressive accumulation of plastic activity, with clear signatures of dimensionality in the failure time divergence exponent approaching to criticality from above. Although structural features can influence where and when failure occurs, a significant component of the stochasticity in failure time appears to be intrinsic to the dynamics approaching failure. 

\begin{acknowledgments}
%\acknowledgement{
We thank Misaki Ozawa, Pushkar Khandare, Swarnendu Maity, and Debargha Sarkar for useful discussions. We acknowledge the National Supercomputing Mission facility (Param Yukti) at the Jawaharlal Nehru Centre for Advanced Scientific Research for computational resources. 
%}
\end{acknowledgments}
%\funding{ S.S. acknowledges support from the Science and Engineering Research Board (Anusandhan National Research Foundation) (India) through the JC Bose Fellowship (Grant No. JBR/2020/000015) and a grant under Scientific and Useful Profound Research Advancement (SUPRA) (Grant No. SPR/2021/000382).}

%\roles{HB and SS conceived the research, HB performed the research, HB and SS analyzed the data and wrote the paper.}

%\data{Data included in the paper are available from the corresponding author upon reasonable request.}

% \nocite{*}
\bibliography{Bibliography}
\bibliographystyle{apsrev4-1}

\clearpage
\newpage
%\onecolumngrid

%\input{SI_body}

\end{document}